\documentclass[aps,prb,twocolumn,superscriptaddress,showpacs,fleqn,floatfix]{revtex4}

\usepackage[dvips]{epsfig}
\usepackage[dvips]{color}
\usepackage{amssymb}

\newcommand{\bacusio}{BaCu$_2$Si$_2$O$_7$}
\newcommand{\kvo}{K$_2$V$_3$O$_8$}
\newcommand{\ham}{\hat{\cal{H}}}
\newcommand{\svec}{\hat{\mathbf{S}}}

\newcommand{\eqnref}[1]{(\ref{#1})}
\newcommand{\vectorprod}[2]{\bigl[\mathbf{#1}\times\mathbf{#2}\bigr]}
\newcommand{\scalarprod}[2]{\bigl(\mathbf{#1}\cdot\mathbf{#2}\bigr)}
\newcommand{\mixedprod}[3]{\left(\mathbf{#1}\cdot\bigl[\mathbf{#2}\times\mathbf{#3}\bigr]\right)}

\begin{document}
\title{Magnetic resonance study of the spin-reorientation transitions in
the quasi-one-dimensional
antiferromagnet \bacusio{}.}

\author{V.N.Glazkov}
\affiliation{P. L. Kapitza Institute for Physical Problems RAS, 117334
 Moscow, Russia}
\affiliation{Commissariat \`a l'Energie Atomique, DSM/DRFMC/SPSMS,
38054 Grenoble Cedex 9, France}

\author{A.I.Smirnov}
\affiliation{P. L. Kapitza Institute for Physical Problems RAS, 117334
 Moscow, Russia}

\author{A.Revcolevschi}
\author{G.Dhalenne}
\affiliation{Laboratoire de Physico-Chimie de l'Etat Solide, Universit\'e Paris-Sud,
91405 Orsay Cedex, France}

\date{\today}

\begin{abstract}
A quasi-one dimensional antiferromagnet with a strong reduction of
the ordered spin component, \bacusio,{} is studied by the magnetic
resonance technique in a wide field and frequency range. Besides of
conventional spin-flop transition at the magnetic field parallel to
the easy axis of spin ordering, magnetic resonance spectra indicate
additional spin-reorientation transitions in all three principal
orientations of magnetic field. At these additional transitions the
spins rotate in the plane perpendicular to the magnetic field keeping
the mutual arrangement of ordered spin components. The observed
magnetic resonance spectra and spin-reorientation phase transitions
are quantitatively described by a model including the anisotropy of
transverse susceptibility with respect to the order parameter
orientation. The anisotropy of the transverse susceptibility and the
strong reduction of the anisotropy energy due to the quantum spin
fluctuations are proposed to be the reason of the spin reorientations
which are observed.
\end{abstract}

\maketitle

\section{Introduction}

Low dimensional spin systems have been actively studied during the
last few decades. This non-vanishing interest is due to the
increasing role of quantum fluctuations in these magnets, which can
lead to the formation of non-trivial ground sates. One-dimensional
antiferromagnetic spin chains are known to remain in a
quantum-disordered state down to zero temperature. However, a weak
inter-chain interaction makes the formation of  Neel order again
possible.

The formation of Neel order in these quasi-one-dimensional magnets is
characterized by two features: Firstly, the transition temperature
$T_N$ is much smaller than the temperature corresponding to the
in-chain exchange integral. Secondly, the sublattice magnetization
(average spin value per site) is strongly reduced with respect to the
nominal value.

The recently studied oxide \bacusio{} is a good example of a spin
S=1/2 quasi-one-dimensional antiferromagnet. The exchange integral
values determined by inelastic neutron scattering \cite{Kenzelmann}
are $J$=24.1meV for the in-chain interaction and $J_a$=-0.46meV,
$J_b$=0.20meV, $J_{[110]}$=0.076meV for different inter-chain
interactions. The ordering temperature is $T_N$=9.2K. Susceptibility
data \cite{tsukada:2sf} reveal that the $c$ axis is the easy axis of
the spin ordering. The ordered magnetic moments at lattice sites were
found to be equal to 0.15$\mu_B$ at zero magnetic
field.\cite{Kenzelmann,zheludev:struct} Thus a strong spin reduction
by zero point fluctuations is present in this material.

The interest for the magnetic order observed in \bacusio{} was
enhanced by the observation of unusual spin-reorientation
transitions: when the magnetic field is applied along the easy
anisotropy axis
 two spin-reorientation transitions were observed
\cite{tsukada:2sf} at the fields $H_{c1}$=20~kOe and $H_{c2}$=49~kOe.
Later, another phase transition at $H_{c3}$=78~kOe was reported for
$\mathbf{H}||b$,\cite{poirier:sf3}  i.e. when the field is
perpendicular to the easy axis. The magnetic structure at
$\mathbf{H}||c$ was identified with the help of elastic neutron
scattering experiments:\cite{zheludev:struct} at low fields,
$H<H_{c1}$, the local magnetic moments are aligned
antiferromagnetically along the $c$-axis. At the first transition (at
$H_{c1}$) they  rotate to the $(ab)$ plane and align along the $b$
axis. Finally, at the
second transition field, $H_{c2}$, the antiferromagnetic vector is
rotated within the $(ab)$ plane towards the $a$ axis. Neglecting a
small spin canting in the high-field phases, the exchange spin
structure (mutually parallel or antiparallel orientation of the local
magnetic moments) remains the same in all three phases.

The observation of the spin-reorientation transitions at $H_{c2}$ and
$H_{c3}$ is surprising for a supposedly  collinear antiferromagnet.
Usually a collinear easy-axis antiferromagnet exhibits only a
spin-reorientation known as a spin-flop when the magnetic field is
applied parallel to the easy axis. At the spin-flop point, the
antiferromagnetic order parameter rotates from the easy axis to a
perpendicular direction, and the loss in anisotropy energy is
compensated by the gain in magnetization energy due to the large
transverse susceptibility.

A unique spin-reorientation transition when the field is applied
perpendicular to the easy axis of a collinear antiferromagnet, was
reported earlier for \kvo{}.\cite{kvo} Here, a weak ferromagnetic
moment arises if the antiferromagnetic order parameter is tilted
apart from the easy axis aligned along the high order symmetry axis.
In this case the gain in Zeeman energy is due to the appearance of a
weak ferromagnetic moment. This is not the case of \bacusio{}: no
signs of a ferromagnetic magnetization in high-field phases were
present, and, moreover, the identified antiferromagnetic structure is
not compatible with weak ferromagnetism. \cite{glazkov}

In the theoretical model of Ref.\onlinecite{glazkov}  two additional
phase transitions observed in \bacusio{} were explained by a
relatively strong dependence of the transverse susceptibility on the
direction of the order parameter. The resulting anisotropic part of
the Zeeman energy exceeds the change in energy of magnetic anisotropy
above the corresponding critical fields. However, the origin of this
anisotropy, as well as the reasons for the increase of its influence,
are still not understood.

The magnetic resonance technique enables one to explore  low-energy
dynamics of magnetic crystals by sensitive and precise measurement of
the resonance frequency and polarization of $\mathbf{q}=0$
excitations. At spin-reorientation transitions the local ordered
moments rotate by a noticeable angle,  and, at the exact field value,
this rotation costs no energy. Thus, the corresponding mode of the
spin resonance is softening at the transition field and its frequency
turns to zero. Therefore the observation of the magnetic resonance
soft modes appears to be a powerful tool to study spin-reorientation
transitions.

We present here a detailed magnetic resonance  study of \bacusio{},
which is essentially complementary to earlier data of Refs.
\onlinecite{jpsj-afmr,hayn-afmr}. We have observed the soft modes of
the antiferromagnetic resonance (AFMR), marking the mentioned
spin-reorientation transitions for $\mathbf{H}||c$ and
$\mathbf{H}||b$. Besides, we report on the spin-reorientation
transition for $\mathbf{H}||a$, at a magnetic field of 110 kOe, which
has not been observed before. We present a model based on the
macroscopic approach to the spin dynamics of an antiferromagnet which
describes quantitatively the observed AFMR spectra and the
field-induced phase transitions. The model includes the anisotropy of
the transverse susceptibility with respect to the order parameter
orientation. We discuss the possible origin of this anisotropy and
the possible increase of its influence because of the strong
reduction of the ordered moments by quantum fluctuations.

\section{Experimental technique and samples preparation.}

Magnetic resonance was studied using a set of home-made ESR
spectrometers with transmission type resonators covering the
frequency range 9--110 GHz, equipped with  He-cooled cryomagnets for
8 and 14T. Magnetic resonance lines were recorded by measurements of
the field-dependences of the microwave power transmitted through the
resonator with the sample.

Single crystalline samples of \bacusio{}, several centimeters long,
were grown by the floating zone technique \cite{poirier:sf3}, from
which samples of about 2x2x3 mm$^3$ were cut and used for our
microwave experiments.  Crystals were oriented by a conventional
x-ray diffraction method.  Because the lattice parameters for $a$ and
$c$ directions are very close ($a$=6.862\AA{} $b$=13.178\AA,
$c$=6.891\AA{} \cite{cryst-struct}), only $b$ axis can be easily
unambiguously identified. The axis $a$ and $c$ were distinguished by
the behaviour in the magnetic field using the data of
Ref.\onlinecite{tsukada:2sf}, where $c$ axis was identified as an
easy axis of magnetic ordering.

\section{Experimental results}
\begin{figure}
    \centering
    \epsfig{file=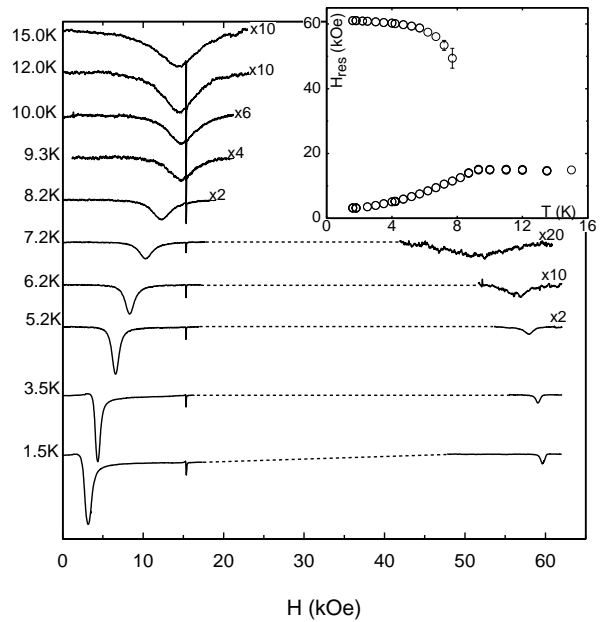, width=0.9\columnwidth, clip=}
    \caption{Temperature dependence of the ESR absorption spectrum at $f=44.04$GHz
    and  $\mathbf{H}||b$.
      Dashed lines are guides to the eye connecting separately measured low-field
      and high-field parts of the ESR absorption spectrum.
      High-field parts of the  T=5.2 K, 6.2 K and 7.2 K curves
      are Y-stretched with the corresponding coefficient. The narrow
      absorption line at
      $H=$15.7 kOe
      is a DPPH mark ($g=2.0$). Insert: temperature dependence of the
      two resonance fields.}
    \label{fig:line(t)b}
\end{figure}
\begin{figure}
    \centering
    \epsfig{file=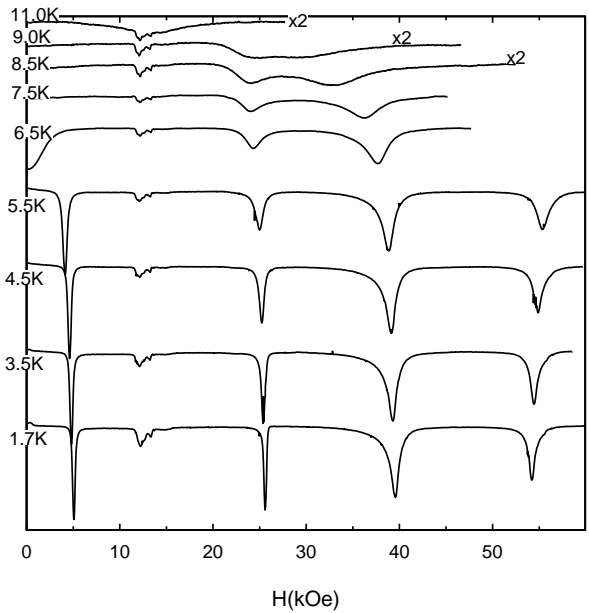, width=0.9\columnwidth, clip=}
    \caption{Temperature dependence of the ESR absorption spectrum at $f=38.25$GHz
    and with
    $\mathbf{H}||c$.}
    \label{fig:line(t)c}
\end{figure}
\begin{figure}
    \centering
    \epsfig{file=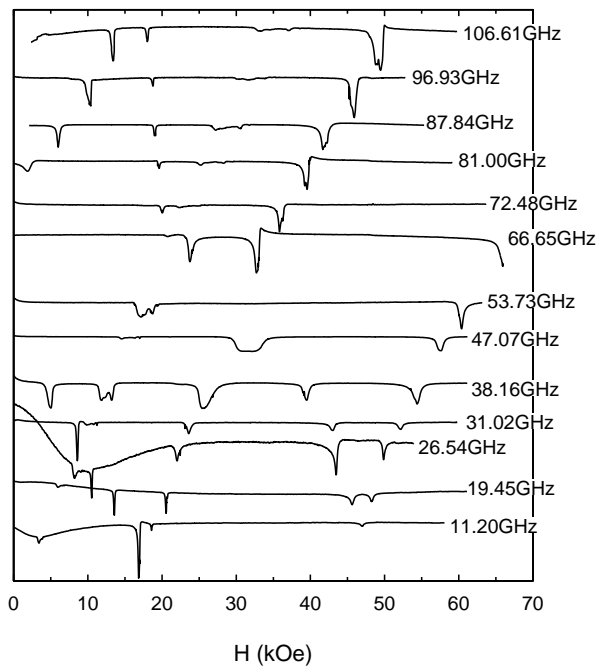, width=0.9\columnwidth, clip=}
    \caption{ESR absorption spectrum at different microwave frequencies.
    $\mathbf{H}||c$, T=1.5K.}
    \label{fig:line(f)c}
\end{figure}
\begin{figure}
    \centering
    \epsfig{file=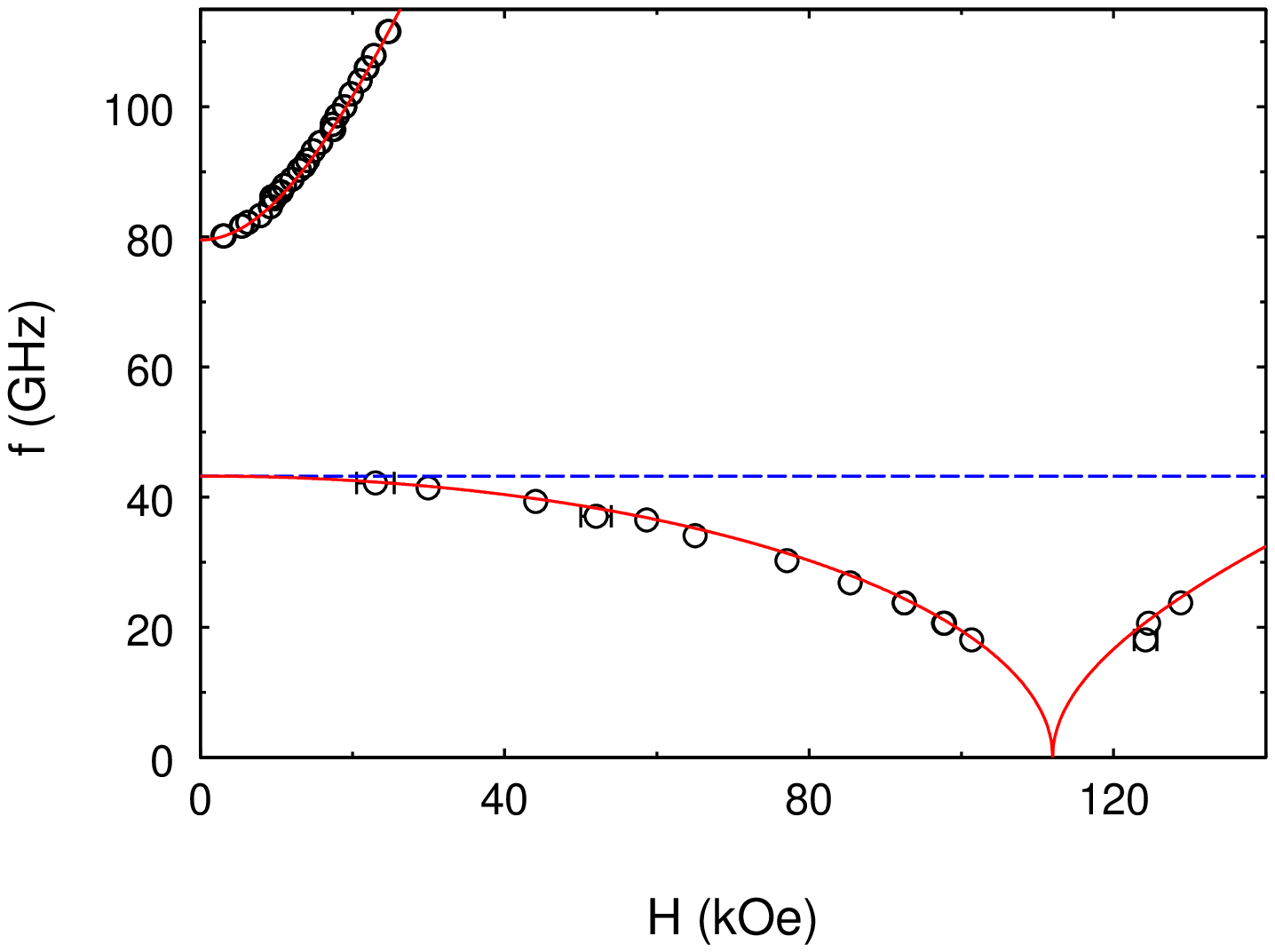, width=0.9\columnwidth, clip=}
    \caption{(color online)
        Field dependence of the antiferromagnetic resonance frequency at
         $\mathbf{H}||a$ and $T=$1.5~K.
        Circles represent experimental points of resonance absorption and
        solid lines --- theory (see text).
            Dashed line is the $f(H)$ dependence calculated
            for a two sublattices antiferromagnet with the same
            zero-field gaps.}
    \label{fig:spectr-a}
\end{figure}
\begin{figure}
    \centering
    \epsfig{file=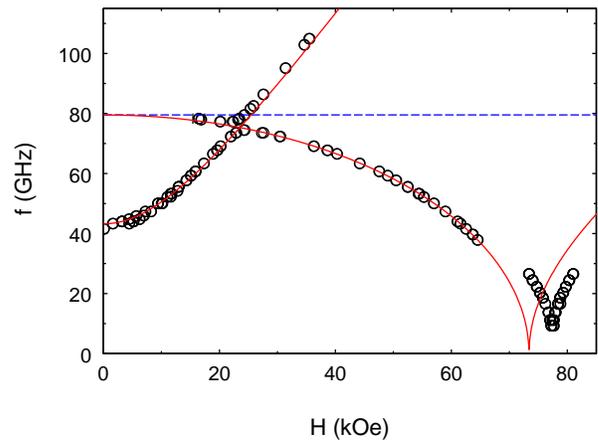, width=0.9\columnwidth, clip=}
    \caption{(color online)
Field dependence of the antiferromagnetic resonance frequency at
$\mathbf{H}||b$, and $T=$1.5K. Circles are experimental points and
solid line is theoretical calculation (see text).
    Dashed lines is the $f(H)$  dependence of a two
    sublattices antiferromagnet with the same
    zero-field gaps.}    \label{fig:spectr-b}
\end{figure}
\begin{figure}
    \centering
    \epsfig{file=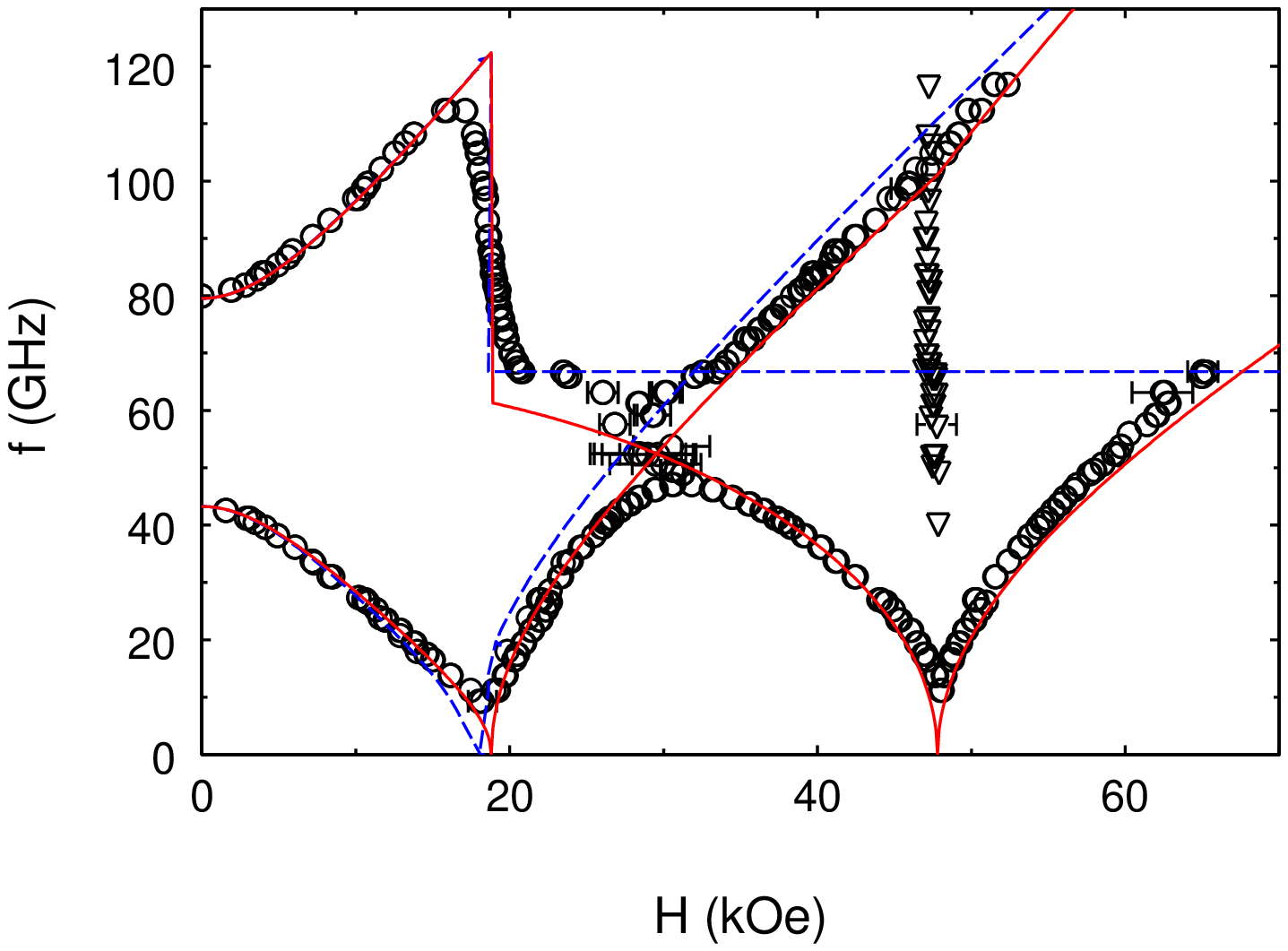, width=0.9\columnwidth, clip=}
    \caption{(color online)
Field dependence of the antiferromagnetic resonance frequency at
    $\mathbf{H}||c$, $T=$1.5~K. Circles represent experimental points of
    resonance absorption, triangles --- non-resonance response at the
    spin-reorientation  transition, solid line --- theory (see text).
Dashed lines is the $f(H)$ dependence calculated for a two
sublattices antiferromagnet with the same
            zero-field gaps.}        \label{fig:spectr-c}
\end{figure}

The temperature evolution of the resonance absorption spectrum at
$\mathbf{H}||b$ is shown at Figure \ref{fig:line(t)b}. At $T\geq 9.3K$,
a broad resonance line with a g-factor slightly above 2.0 is observed.
The absorption spectrum changes when crossing the transition temperature
$T_N=9.2K$. The maximum of absorption shifts to lower fields and another
absorption line appears at a higher field. The temperature dependence of
the resonance field for both resonance lines is shown in the insert of
the Fig. \ref{fig:line(t)b}. The temperature at which the low-field
absorption line begins to shift corresponds well to  $T_N$. When the
field is applied along the easy axis (see Fig. \ref{fig:line(t)c}) a
multi-component absorption spectrum develops below $T_N$.

An additional absorption signal of irregular shape is clearly visible
on Fig. \ref{fig:line(t)c} at a field close to 13~kOe. Its intensity
increases with decreasing temperature and its position is temperature
independent. Probably it is due to a paramagnetic impurity or to a
small amount of another phase.

Magnetic resonance absorption curves, taken at different frequencies,
are illustrated in Fig. \ref{fig:line(f)c}. The frequency-field
dependences $f(H)$ are shown at Figures
\ref{fig:spectr-a}---\ref{fig:spectr-c}. Two modes of resonance
absorption with  zero-field gaps of approximately 42 and 80 GHz are
observed for all three orientations of the magnetic field. The low-field
spectra are in good agreement with the $f(H)$ dependences of a
conventional two-sublattice antiferromagnet, \cite{Kubo-afmr} which are
represented by dashed lines in
Figs.~\ref{fig:spectr-a}---\ref{fig:spectr-c}: when the magnetic field
is applied parallel to the easy axis a characteristic spin-flop
transformation of the spectrum, with a vertical frequency drop, occurs
at $H_{c1}=(18.8\pm0.8)$kOe. For the other directions, in the low-field
range $H\leq$20~kOe there is one mode with a strong field dependence of
the resonance frequency and one mode with an approximately constant
frequency. At $\mathbf{H}||a$ the resonance mode with a strong field
dependence starts from the larger zero-field gap, while at
$\mathbf{H}||b$ the larger gap corresponds to a mode with a weak field
dependence of the resonance frequency. This low-field behavior allows to
identify the $a$ axis as a hard axis of antiferromagnetic ordering and
$b$ axis --- as a middle axis of spin ordering according to
Ref.~\onlinecite{Kubo-afmr}.

However, upon increasing the magnetic field, an additional softening
of one of the two modes occurs. An additional softening occurs for
the easy-axis orientation of the applied field ($\mathbf{H}||c$) at
the field $H_{c2}=(47.8\pm1.0)$kOe, for the $\mathbf{H}||b$ at the
field $H_{c3}=(77.4\pm0.8)$kOe, and for the  $\mathbf{H}||a$
orientation at the field $H_{c4}=(110\pm5)$kOe.

The values  of the spin-reorientation fields  $H_{c2}$ and $H_{c3}$
are in good agreement with that  reported in
Refs.~\onlinecite{tsukada:2sf,poirier:sf3}. Softening of the AFMR
modes at these fields was reported also in
Refs.~\onlinecite{hayn-afmr,jpsj-afmr} (note that  the orientations
$\mathbf{H}||a$ and $\mathbf{H}||b$ are mixed in Fig.12 of
Ref.\onlinecite{jpsj-afmr}). The last spin-reorientation transition,
observed at $\mathbf{H}||a$ ($H_{c4}$), was not reported before.

\section{Discussion.}

\subsection{Rotations of the order parameter.}

The softening of spin-resonance modes indicates spin-reorientation
transitions, since at the transition field the rotation of the spins
costs no energy. Comparison of the observed antiferromagnetic
resonance frequency-field dependences with the results of the
conventional two-sublattice model \cite{Kubo-afmr} is shown in
Figs.~\ref{fig:spectr-a}---\ref{fig:spectr-c}. This matching allows
one to draw simple conclusions on the directions of the order
parameter rotation at the observed spin-reorientation transitions.

The standard model of the two-sublattices antiferromagnet includes the
anisotropy energy $a_1l_x^2+a_2l_y^2$ ($l_\alpha$ are order parameter
components) and the Zeeman energy resulting from the transverse
susceptibility. This model yields only the spin-flop transition at the
magnetic field applied parallel to the easy axis of spin ordering. At
the spin-flop point the spins rotate from the easy axis to the
perpendicular direction. There are 8 magnetic ions in the unit cell of
\bacusio{}, therefore this compound should be a multi-sublattice
antiferromagnet. Nevertheless, the low-frequency dynamics of the
collinear antiferromagnet in a {\em weak} magnetic field is described by
a universal equation which does not depend on the number of
sublattices.\cite{Andreev-Marchenko} Therefore, at low fields, i.e.
reasonably below  $H_{c2,c3,c4}$, one can use the classification of the
AFMR modes in terms of the order parameter oscillations in a
two-sublattice antiferromagnet\cite{Kubo-afmr}.

At the field $H_{c1}$ the magnetic resonance mode starting from the
lower gap is softening.  In the two-sublattice model this mode
corresponds to the oscillations of the order parameter in the
direction from the easy axis towards the second-easy axis. Softening
of this mode means the free rotation of the order parameter from easy
axis to the second-easy axis, as at the conventional spin-flop
transition. Further, the data presented on Figs.4-6 indicate that the
AFMR softening observed at the fields $H_{c2,c3,c4}$ corresponds to
the field-independent resonance modes of the two-sublattice
antiferromagnet. These field-independent modes are oscillations of
the order parameter in the plane which is perpendicular to the
magnetic field. Thus, the softening of the AFMR modes at the fields
$H_{c2,c3,c4}$ reveal the spin rotation around the the field
direction. These conclusion is in agreement with the change of the
magnetic structure at the transition field $H_{c2}$, derived from
elastic neutron scattering.\cite{zheludev:struct} The above
identification of resonance modes indicates that when $\mathbf{H}||a$
and $\mathbf{H}||b$ the spin reorientations correspond to the
rotation of the order parameter from the easy axis $c$ to the
direction of $b$ and $a$ axes, correspondingly.

\subsection{Antiferromagnetic resonance theory.}

The comparison of our results with the simple two-sublattice model
allows to draw the above qualitative conclusions about the spin
rotations at the observed phase transitions. Below we give a
quantitative description of the observed antiferromagnetic resonance
frequencies. While obeying the two sublattice model in low fields,
the multisublattice antiferromagnet like \bacusio{} may, in
principle, change the exchange spin structure in high fields of about
the exchange field value. Therefore we should check at first, whether
the observed spin-reorientation transitions conserve the exchange
spin structure (i.e. the mutual alignment of the ordered spin
components).

The field range of the present experiment is well below the value of
the main exchange field $H_E\sim J/\mu_B\sim 4\cdot10^3$kOe derived
from the in-chain exchange integral $J=$24.1meV. In the same time,
the interchain exchange integrals may be of the order of the applied
fields. However, the change of the low-field exchange structure at
the fields $H_{c1,c2,c3,c4}$ may be excluded by the following
experimental observations. At first, the elastic neutron scattering
experiment \cite{zheludev:struct} performed at the easy-axis
orientation of the applied field has revealed that the main component
of the antiferromagnetic order parameter corresponds to the same
mutual alignment of spins in all three phases. Second, as described
above,  we observe at the fields $H_{c1,c2,c3,c4}$ the softening of
the lowest resonance modes with the gaps of the anisotropic nature,
analogous to the modes of  a two sublattice antiferromagnet, while at
the change of the exchange structure the modes of the exchange origin
should be softened. The exchange modes of multisublattice
antiferromagnets \cite{ExchangeModes}, corresponding to the
oscillations of the mutual orientations of the ordered moments have
usually much higher frequencies, of the order of $J/h$, while the
frequencies of lowest $q=0$ spin waves are zero in the absence of
anisotropy. The high-frequency ESR measurements \cite{jpsj-afmr} have
found resonance mode at a frequency of about 400 GHz in the field
range below 150 kOe which probably softens in a field near 250 kOe.
This is clearly one of the exchange modes. We consider it as the
lowest exchange mode.  The magnetic field value $H^*=$250kOe can be
taken as a measure of the exchange field corrected for the
low-dimensionality. The field range of the present experiment is at
least two times smaller than $H^*$.

Thus, we suppose in the further analysis, that the zero-field exchange
structure remains the same in the whole field range of the described
experiment. Under this condition, a macroscopic (hydrodynamic) approach
to the spin dynamics \cite{Andreev-Marchenko} is valid. As a first step
of this approach we write down the potential energy of the
antiferromagnet as a function of  the orientation of the order parameter
and magnetic field. The potential energy expansion includes
usually the anisotropy energy and Zeeman energy. These terms are
responsible for the conventional spin-flop transition in the easy-axis
orientation of the magnetic field. To include additional
spin-reorientation transitions we phenomenologically add the higher
order invariants allowed by the crystal symmetry ($D_{2h}^{16}$) and by
the spin-order parameter. The eight-sublattice order parameter was found in
the experiment\cite{zheludev:struct} and has the form of ${\bf l}_6$ in
terms of Ref.~\onlinecite{glazkov}. This structure is not compatible
with weak ferromagnetism.\cite{glazkov} Thus, the potential energy
contains only terms of even power of $l_{\alpha}$ and $H_{\alpha}$. As
shown in Ref.~\onlinecite{glazkov}, terms quadratic on both $l_{\alpha}$
and $H_{\alpha}$ should be added to explain additional transitions.

We consider in zero-temperature limit the following form of the
potential energy (here the exchange part of the transverse magnetic
susceptibility is set to unity, ${\bf l}$ is the unit vector in the
direction of the order parameter):
\begin{eqnarray}
U&=&-\frac{1}{2}\vectorprod{l}{H}^2+a_1l_x^2+a_2l_y^2+\xi_1\scalarprod{H}{l}H_xl_x
+\nonumber\\
&&+\xi_2\scalarprod{H}{l}H_yl_y
-(\xi_1+\xi_2)\scalarprod{H}{l}H_zl_z-\nonumber\\
&&-B_1H_x^2(l_y^2-l_z^2)-B_2H_y^2(l_x^2-l_z^2)-B_3H_z^2(l_x^2-l_y^2)+\nonumber\\
&&+C_1H_yH_zl_yl_z+C_2H_xH_zl_xl_z+C_3H_xH_yl_xl_y\label{eqn:energy}
\end{eqnarray}

\noindent the Cartesian coordinates are chosen as $x||a$, $y||b$,
$z||c$. The first three terms are Zeeman energy  and magnetic
anisotropy energy, $a_1>a_2>0$ are anisotropy constants corresponding
to  the observed hierarchy of the anisotropy axes in weak fields.
These terms describe adequately the  observed low-field behaviour
with the characteristic features of a collinear antiferromagnet.
Other terms represent all symmetry-allowed invariants quadratic in
components of $\mathbf{H}$ and $\mathbf{l}$. The three invariants
with $\xi_i$ coefficients are of exchange-relativistic origin: they
contain the scalar product $\scalarprod{l}{H}$, which is invariant
under simultaneous rotation of the order parameter and the magnetic
field. Other terms are of relativistic (spin-orbital) nature, they
are not invariant under simultaneous rotations of magnetic vectors.

Equation \eqnref{eqn:energy} is written just phenomenologically, we
do not consider the  hierarchy of the relativistic terms. Potential
energy \eqnref{eqn:energy} differs from that  of
Ref.~\onlinecite{glazkov}, where only two exchange-relativistic terms
$\mathbf{H}^2l_x^2$ and $\mathbf{H}^2l_y^2$ were taken into account
instead of the six terms with the $B_i$ and $C_i$ coefficients in
Eqn.~\eqnref{eqn:energy}. Other terms were neglected because they
should be smaller than exchange-relativistic terms. However, the
model of Ref.~\onlinecite{glazkov} does not describe the
spin-reorientation transition at $\mathbf{H}||a$ ($H_{c4}$). We show
below in sec. C, that the coefficients $B_i$, $C_i$ are of the same
order in spin-orbital coupling.

Minimization of the energy \eqnref{eqn:energy} in the exact
orientations of the magnetic field yields for the fields of
spin-reorientation transitions:

\begin{eqnarray}
\mathbf{H}||z,~\mathbf{l}||z\leftrightarrow\mathbf{l}||y&:&H_{c1}^2=\frac{2a_2}{1-2(\xi_1+\xi_2)-2B_3}\label{eqn:hc1}\\
\mathbf{H}||z,~\mathbf{l}||y\leftrightarrow\mathbf{l}||x&:&H_{c2}^2=\frac{a_1-a_2}{2B_3}\label{eqn:hc2}\\
\mathbf{H}||y,~\mathbf{l}||z\leftrightarrow\mathbf{l}||x&:&H_{c3}^2=\frac{a_1}{2B_2}\label{eqn:hc3}\\
\mathbf{H}||x,~\mathbf{l}||z\leftrightarrow\mathbf{l}||y&:&H_{c4}^2=\frac{a_2}{2B_1}\label{eqn:hc4}
\end{eqnarray}

The transition at $H_{c1}$ is the normal spin-flop transition, it
occurs when the gain in magnetization energy due to the large
transverse susceptibility exceeds the loss in anisotropy energy.
Other transitions result from the gain in the magnetization energy
due to the small difference of the transverse susceptibility in the
phases with different orientations of the  antiferromagnetic vector.
Directions of the order parameter rotation at these transitions are
in agreement with the results of previous subsection.

The dynamics of the collinear antiferromagnet in zero temperature
limit is described \cite{Andreev-Marchenko} by the equation:

\begin{equation}
\frac{\partial}{\partial t}\vectorprod{l}{\dot{l}}=\gamma^2\vectorprod{l}{H_l}
+2\gamma\scalarprod{l}{H}\mathbf{\dot{l}}
\end{equation}

\noindent here $\mathbf{H_l}=-\partial U/\partial\mathbf{l}$,
exchange part of the transverse susceptibility is set to unity as
before. Linearizing equation (6) near the equilibrium spin
configuration and solving for eigenfrequencies, we obtain the
following expressions for the AFMR frequency-field dependences at the
exact orientations of the magnetic field:\\ \noindent for
$\mathbf{H}||x$ (hard axis of magnetization):\\ \noindent$H<H_{c4}$:
\begin{eqnarray}
\frac{\nu_1^2}{\gamma^2}&=&\bigl(1+2\xi_1\bigr)H^2+2a_1-a_2\left(\frac{H}{H_{c4}}\right)^2\label{eqn:afmr-theory-start}\\
\frac{\nu_2^2}{\gamma^2}&=&2a_2\left[1-\left(\frac{H}{H_{c4}}\right)^2\right]
\end{eqnarray}
\noindent$H>H_{c4}$:
\begin{eqnarray}
\frac{\nu_1^2}{\gamma^2}&=&\bigl(1+2\xi_1\bigr)H^2+2\bigl(a_1-a_2\bigr)+a_2\left(\frac{H}{H_{c4}}\right)^2\\
\frac{\nu_2^2}{\gamma^2}&=&2a_2\left[\left(\frac{H}{H_{c4}}\right)^2-1\right]
\end{eqnarray}

\noindent for $\mathbf{H}||y$ (second-easy axis of magnetization):\\
\noindent$H<H_{c3}$:
\begin{eqnarray}
\frac{\nu_1^2}{\gamma^2}&=&2a_1\left[1-\left(\frac{H}{H_{c3}}\right)^2\right]\\
\frac{\nu_2^2}{\gamma^2}&=&\bigl(1+2\xi_2\bigr)H^2+2a_2-a_1\left(\frac{H}{H_{c3}}\right)^2
\end{eqnarray}
\noindent$H>H_{c3}$:
\begin{eqnarray}
\frac{\nu_1^2}{\gamma^2}&=&2a_1\left[\left(\frac{H}{H_{c3}}\right)^2-1\right]\\
\frac{\nu_2^2}{\gamma^2}&=&\bigl(1+2\xi_2\bigr)H^2-2\bigl(a_1-a_2\bigr)+a_1\left(\frac{H}{H_{c3}}\right)^2
\end{eqnarray}

\noindent for $\mathbf{H}||z$ (easy axis of magnetization):\\
\noindent$H<H_{c1}$:
\begin{eqnarray}
\frac{\nu_{1,2}^2}{\gamma^2}&=&\bigl(1+2(\xi_1+\xi_2)\bigr)H^2+a_1+a_2\pm\nonumber\\
&\pm&\Biggl[4\left(a_1+a_2-\frac{1}{2}\left(\frac{a_1-a_2}{H_{c2}}\right)^2\right)H^2+(a_1-a_2)^2+ \nonumber\\
&&+\left(\left(\frac{a_1-a_2}{H_{c2}^2}\right)^2+8(\xi_1+\xi_2)\right)H^4\Biggr]^{1/2}
\end{eqnarray}
\noindent$H_{c1}<H<H_{c2}:$
\begin{eqnarray}
\frac{\nu_1^2}{\gamma^2}&=&2(a_1-a_2)\left[1-\left(\frac{H}{H_{c2}}\right)^2\right]\\
\frac{\nu_2^2}{\gamma^2}&=&\bigl(1-2(\xi_1+\xi_2)\bigr)H^2-2a_2-(a_1-a_2)\left(\frac{H}{H_{c2}}\right)^2
\end{eqnarray}
\noindent$H>H_{c2}$:
\begin{eqnarray}
\frac{\nu_1^2}{\gamma^2}&=&2(a_1-a_2)\left[\left(\frac{H}{H_{c2}}\right)^2-1\right]\\
\frac{\nu_2^2}{\gamma^2}&=&\bigl(1-2(\xi_1+\xi_2)\bigr)H^2-2a_1+(a_1-a_2)\left(\frac{H}{H_{c2}}\right)^2\label{eqn:afmr-theory-stop}
\end{eqnarray}

\noindent For $\xi_{1,2}=0$, $B_{1,2,3}=0$ the transition fields
$H_{c2,c3,c4}$ turn to infinity and the frequencies correspond to the
well known case of a collinear antiferromagnet.
\cite{Andreev-Marchenko,Kubo-afmr}

Equations
\eqnref{eqn:afmr-theory-start}---\eqnref{eqn:afmr-theory-stop}
describe experimental data with 8  fitting parameters: the
coefficients $a_{1,2}$ determine the zero-field gaps, transition
fields $H_{c2,c3,c4}$ are found as  fields of softening of different
resonance modes, $\gamma$ and $\xi_{1,2}$ affect the slope of the
magnetic resonance branches. Parameters $B_i$ of the potential energy
\eqnref{eqn:energy} can be then recalculated by the
Eqns.~\eqnref{eqn:hc1}---\eqnref{eqn:hc2}. Comparison with the
experiment is presented in the Figures
\ref{fig:spectr-a}---\ref{fig:spectr-c}, and the fit is very good.
The deviations from the calculated curves  near the intersections  of
the two resonance branches are due to the small misorientation of the
sample. The discrepancy of the low-frequency $\mathbf{H}||b$
experimental data with the theoretical curve  is most probably due to
a small tilting of the sample during these measurements.

Values of the best fit parameters are: $\gamma=2.82$GHz/kOe,
$a_1=400$kOe$^2$, $a_2=118$kOe$^2$, $B_1$=0.0047, $B_2=0.0370$,
$B_3=0.0614$, $\xi_1=0.135$, $\xi_2=-0.03$. Corresponding transition
fields  are: $H_{c1}=18.0$kOe, $H_{c2}=47.8$kOe, $H_{c3}=73.4$kOe and
$H_{c4}=110$kOe

The $C_i$ parameters of Eqn.~\eqnref{eqn:energy} do not affect the
resonance frequencies at the exact orientations of the magnetic field
because the corresponding terms turn to zero.

\subsection{Microscopic approach.}
\begin{figure}
    \centering
    \epsfig{file=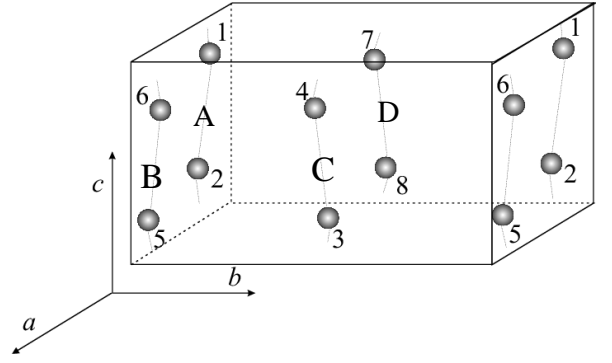, width=0.9\columnwidth, clip=}
    \caption{Unit cell of \bacusio{}. Notation of the ion positions and unequivalent chains.}
    \label{fig:ucell}
\end{figure}

In the above consideration we have described the observed
frequency-field dependences of the antiferromagnetic resonance modes
assuming that the transverse magnetic susceptibility is essentially
anisotropic with respect to the orientation of the order parameter.
The additional transitions occur in this model as a result of the
competition between the loss in the anisotropy energy and of the gain
in  Zeeman energy,  due to the increase of transverse magnetization
at a rotation of the order parameter. For an ordinary antiferromagnet
these terms become comparable at high magnetic fields, of the order
of the exchange field. Hence, in order to  explain  the observed
moderate values of the transition fields, one has to imply (see
Eqns.~\eqnref{eqn:hc2}---\eqnref{eqn:hc4}) that either the $B_i$
coefficients are unusually large, or  the anisotropy parameters $a_i$
are particularly small.

The qualitatitive microscopic analysis given below demonstrates that
the spin reduction affects differently anisotropy constants and the
anisotropy of the transverse susceptibility and, therefore, is a
likely reason of unusual spin-reorientation transitions. We start
from the microscopic Hamiltonian and then proceed with a classical
approximation, considering spins as vectors of length $S$. This
classical approach cannot account for all possible effects of spin
reduction, e.g. this consideration  neglects the most probable
anisotropy of the quantum spin fluctuations, which could be another
reason for the anisotropy of the transverse susceptibility and
contribute also to the induction of the described spin reorientation
transitions. Our intention here is only to show qualitatively that
effect of spin reduction is different for different terms in the
energy even in classical approximation.

In the model described below we consider only in-chain
Dzyaloshinskii-Moriya (DM) interaction and $g$-tensor anisotropy.
Both of them originate from the relativistic spin-orbital interaction
and are of the same order in spin-orbital coupling:\cite{Moriya}
$D/J\sim\delta g/g$. As before, we use Cartesian coordinates related
to the orthorhombic unit cell: $x||a$, $y||b$, $z||c$.

As a first step, we consider the one-dimensional spin-chain running
along $c$ axis. Note, that there are four inequivalent chains passing
through the unit cell (see Figure \ref{fig:ucell}). Magnetic ions
within a chain are connected by the screw axis of symmetry $C_2'$
aligned along the $c$ axis.

The in-chain Hamiltonian can be written as
\begin{equation}
\ham=\sum_i\left\{\scalarprod{\svec_i}{\svec_{i+1}}+\mixedprod{D_i}{\svec_i}{\svec_{i+1}}+\mathbf{h}\widetilde{g_i}\svec_i\right\}
\label{eqn:ham-start}
\end{equation}

\noindent Here  all interactions are normalized by the in-chain
exchange integral, which is supposed to be antiferromagnetic. The
magnetic field is normalised by the relation
$\mathbf{h}=(\overline{g}\mu_B\mathbf{H}/J)$. The  $g$-tensor is also
normalised by the average $g$-factor:
($\widetilde{g}/\overline{g})\leftrightarrow \widetilde{g}$. Thus,
for the renormalized $g$-tensor $Tr(\widetilde{g})=3$, and in the
isotropic case $g_{\alpha\beta}=\delta_{\alpha\beta}$.

The screw axis puts constraints on the components of the
Dzyaloshinskii-Moriya interaction vector $\mathbf{D}_i$ and on the
components of the $g$-tensor $\widetilde{g_i}$; they can be written
as a sum of staggered and uniform parts.

\begin{eqnarray}
\mathbf{D}_i&=&(-1)^i
\left(
\begin{array}{c}
\alpha\\ \beta\\0
\end{array}
\right)
+
\left(
\begin{array}{c}
0\\0\\ \gamma
\end{array}
\right)\\
\widetilde{g_i}&=&
\left(
\begin{array}{ccc}
g_{xx}&g_{xy}&0\\
g_{xy}&g_{yy}&0\\
0&0&g_{zz}
\end{array}
\right)
+(-1)^i
\left(
\begin{array}{ccc}
0&0&g_{xz}\\
0&0&g_{yz}\\
g_{xz}&g_{yz}&0
\end{array}
\right)
\end{eqnarray}

The direction of the DM-vector is determined by the geometry of superexchange bonds:
for the case of a simple copper-oxygen-copper superexchange path in
\bacusio{} it should be orthogonal to the plane
of the copper-oxygen
bonds. This orientation results  in the relation of the DM vector
components $|D_x|:|D_y|:|D_z|=1.70:1.00:0.16$. Here the crystal
structure parameters of Ref.~\onlinecite{cryst-struct} are used. The
above relation shows that the staggered component of DM vector
dominates in the case of \bacusio{}.

The presence of the uniform component of the Dzyaloshinskii-Moriya
interaction may cause an incommensurate helicoidal spin structure. In
order to find the wavevector of the ground state we consider the
Fourier-transformed spin vectors
$\mathbf{S}_i=\sum_{\mathbf{k}}\mathbf{S}_{\mathbf{k}}e^{\imath(\mathbf{k}
\cdot\mathbf{r}_i)}$ and minimize numerically the Hamiltonian
\eqnref{eqn:ham-start} over the wavevectors $\mathbf{k}$. Thus we
have found that for a ratio between the uniform and staggered DM
vector components less then 0.500, the commensurate state is favored.
For the \bacusio{} structure, the ratio of the uniform $z$-component
of the DM vector to the staggered $xy$-component equals to 0.08,
which certainly corresponds to the commensurate phase.

The commensurate state of a spin chain is a two-sublattice state, it
may be described by the two orthogonal vectors  $\mathbf{S}_0$ and
$\mathbf{S}_\pi$. Spin vector at the $i$-th site of the chain is
expressed via these Fourier components as
$\mathbf{S}_i=\mathbf{S}_0+(-1)^i\mathbf{S}_\pi$. Energy of spin
chain (per spin) can be written as

\begin{equation}
E=\mathbf{S}_0^2-\mathbf{S}_\pi^2+2\mixedprod{S_0}{S_\pi}{D}+\mathbf{h}\widetilde{g_u}\mathbf{S}_0+\mathbf{h}\widetilde{g_s}\mathbf{S}_\pi
\end{equation}
\noindent Here and further on, $\mathbf{D}$ denotes the staggered
component of the DM vector only; $\widetilde{g_u}$ and
$\widetilde{g_s}$ are uniform and staggered components of $g$-tensor,
respectively.

The next step is to integrate over small component $\mathbf{S}_0$ to
obtain the energy as a function of only the antiferromagnetic order
parameter $\mathbf{l}=\mathbf{S}_\pi/|\mathbf{S}_\pi|$ and magnetic
field. This integration is performed for $|\mathbf{S_0}|\ll S$. We
keep all terms with field dependence not higher than $h^2$ and with
coefficients not higher in spin-orbital coupling  than $D^2$. The
condition $|\mathbf{S_0}|\ll S$ effectively implies that our analysis
is valid only in small fields $h\ll h_E=4S$.

The energy of the antiferromagnetically ordered chain per spin is:
\begin{eqnarray}
E&=&-S^2-\frac{1}{8}[(\widetilde{g_u}\mathbf{h})\times\mathbf{l}]^2-\frac{S^2}{2}\vectorprod{l}{D}^2+\nonumber\\
&&+\frac{1}{32}\vectorprod{h}{l}^2\vectorprod{D}{l}^2+\frac{1}{16}\mixedprod{h}{l}{D}^2-\nonumber\\
&&-\frac{1}{8}\mixedprod{h}{l}{D}(\mathbf{h}\widetilde{g_s}\mathbf{l})-\frac{S}{2}(\widetilde{g_u}\mathbf{h})\cdot\vectorprod{l}{D}
\label{eqn:energy-chain}
\end{eqnarray}

\noindent Note, that in zero external field the order parameter
$\mathbf{l}$ obeys the easy plane anisotropy and is confined to the
plane, perpendicular to the staggered component of the
Dzyaloshinskii-Moriya vector $\mathbf{D}$.

To find the total energy per unit cell it is necessary to sum
contributions of all unequivalent chains taking into account
inter-chain exchange interactions and symmetry-implied relations of
Hamiltonian parameters in different chains (denoted as A, B, C and
D):

\begin{eqnarray}
\mathbf{D}_A&=&-\mathbf{D}_B\\
\mathbf{D}_C&=&-\mathbf{D}_D=
\left(
\begin{array}{c}
\alpha_A\\ -\beta_A\\0
\end{array}
\right)
\end{eqnarray}

\begin{eqnarray}
\widetilde{g_u}_A&=&\widetilde{g_u}_B\\
\widetilde{g_s}_A&=&-\widetilde{g_s}_B\\
\widetilde{g_u}_C&=&\widetilde{g_u}_D=
\left(
\begin{array}{ccc}
g_{xx,A}&-g_{xy,A}&0\\
g_{xy,A}&g_{yy,A}&0\\
0&0&g_{zz,A}
\end{array}
\right)\\
\nonumber\\
\widetilde{g_s}_C&=&-\widetilde{g_s}_D=
\left(
\begin{array}{ccc}
0&0&-g_{xz,A}\\
0&0&g_{yz,A}\\
-g_{xz,A}&g_{yz,A}&0
\end{array}
\right)
\end{eqnarray}

To simplify calculations we consider the exactly collinear state
($\mathbf{l}_A||\mathbf{l}_B||\mathbf{l}_C||\mathbf{l}_D$) with
individual antiferromagnetic vectors corresponding to the order
parameter observed in \bacusio{} ($\mathbf{l}_6$ in terms of Ref.\onlinecite{glazkov}):
$\mathbf{l}=\frac{1}{4}(\mathbf{l}_A+\mathbf{l}_B-\mathbf{l}_C-\mathbf{l}_D)$.
This order parameter corresponds to the minimum of the inter-chain
exchange energy. By forcing all chains order parameters to be
collinear we neglect a small contribution to the energy due to the weak canting of the antiferromagnetic structure, which is negligible, as checked by calculations.

As a result we find the following expression for the energy in the
AFM state (per unit cell, additive constants omitted)

\begin{eqnarray}
E&=&-\vectorprod{h}{l}^2+4S^2\alpha^2l_x^2+4S^2\beta^2l_y^2+\nonumber\\
&&+2\left(\sum_{i=x,y,z}(g_{ii}-1)h_il_i\right)\scalarprod{h}{l}+\nonumber\\
&&+\left(\sum_{i=x,y,z}(g_{ii}-1)h_il_i\right)^2+\nonumber\\
&&+g_{xy}^2(h_xl_y+h_yl_x)^2+\frac{1}{4}D^2\vectorprod{h}{l}^2-\nonumber\\
&&-\frac{1}{4}\vectorprod{h}{l}^2(\alpha^2l_x^2+\beta^2l_y^2)+\frac{1}{2}\alpha^2(l_yh_z-h_yl_z)^2+\nonumber\\
&&+\frac{1}{2}\beta^2(l_xh_z-l_zh_x)^2+\alpha g_{yz}(h_z^2l_y^2-h_y^2l_z^2)+\nonumber\\
&&+\beta g_{xz}(h_x^2l_z^2-h_z^2l_x^2)\label{eqn:energy-cell}
\end{eqnarray}

\noindent Note, that the terms which are linear on  the components of
magnetic field $\mathbf{h}$ (see Eqn.~\eqnref{eqn:energy-chain})
exactly compensate each other. All four easy planes of magnetization
of in-chain order parameters intersect along the $z$ axis, marking it
as an easy axis of magnetization in agreement with experimental
observations.

The energy  in \eqnref{eqn:energy-cell} reminds the
phenomenologically written potential energy \eqnref{eqn:energy}. It
includes the field-independent anisotropy of the order parameter,
exchange-relativistic terms similar to the terms with $\xi_{1,2}$
coefficients in \eqnref{eqn:energy}, and the different relativistic
terms quadratic over components of $\mathbf{l}$ and $\mathbf{h}$. The
only essential difference is the presence of a term of fourth order
in the components of $\mathbf{l}$ in Eqn.~\eqnref{eqn:energy-cell}.
We consider its effects below.

Now we point to the direct consequencies of
Eqn.~\eqnref{eqn:energy-cell}.  Anisotropy terms are proportional to
$S^2$, while field-dependent terms do not depend on  $S$. Hence, in
the antiferromagnet with strongly reduced sublattice magnetization
anisotropy terms are strongly reduced in comparison with a
conventional antiferromagnet. Direct comparison of
Eqn.~\eqnref{eqn:energy} and Eqn.~\eqnref{eqn:energy-cell} yields for
the anisotropy constants $a_{1,2}=2S^2D_{x,y}^2/(g^2\mu_B^2)$.
Amplitude of the Dzyaloshinskii-Moriya interaction in the
isostructural germanate BaCu$_2$Ge$_2$O$_7$ determined from the value
of weak ferromagnetic moment is equal to 18K.\cite{bacugeo} Taking
for estimation $D_\alpha=10$K  and  assuming unreduced spin value
$S=1/2$, we obtain for the  value of anisotropy constant
$a\sim3000$kOe$^2$, which is about one order of magnitude larger then
the values determined from the fit of AFMR frequency-field
dependences (400 and 118 kOe$^2$). This reduced anisotropy  may
explain why the spin-reorientations which should be observed in
fields of the order of exchange field are shifted to  lower fields.

Second, the exchange-relativistic terms
$2(g_{ii}-1)h_il_i\scalarprod{l}{h}$  are of the lower order in the
spin-orbital coupling than other terms, quadratic in  components of
$\mathbf{l}$ and $\mathbf{h}$ . Indeed, $(g_{ii}-1)$ has the same
order as $D$, while all the other coefficients are of the order of
$D^2$. This means, in particular, that none of relativistic terms can
be neglected in the Eqn.~\eqnref{eqn:energy}. Since
exchange-relativistic terms $2(g_{ii}-1)h_il_i\scalarprod{l}{h}$ have
coefficients of lower order in the spin-orbital coupling, these terms
could be taken into account even for a standard antiferromagnet:
their effects become comparable with the anisotropy energy in the
field $H_g\sim H_A^{1/4}H_E^{3/4}\ll H_E$.

The fourth order term in \eqnref{eqn:energy-cell} does not affect the
phenomenological analysis of the previous subsection. It can be
decomposed to a sum of the second and fourth order terms
$\Bigl(\mathbf{h}^2-\scalarprod{h}{l}^2\Bigr)\bigl(\alpha^2l_x^2+\beta^2l_y^2\bigr)$.
The fourth order term written in this way does not affect static
properties in the exact orientations of the magnetic field (it is
always zero). Its  dynamic effect is only small changes of the slope
of AFMR branches, but the correction to the slope are dominated by
the $g$-factor anisotropy producing terms of lower  order in
spin-orbital coupling ($D$ instead of $D^2$).

Finally, it is possible to calculate straightforwardly the magnetic
susceptibility in low fields for several orientations of the magnetic
field and of the order parameter. Taking an isotropic $g$-tensor and
keeping in mind that  from the exchange bond geometry $\alpha>\beta$,
one can ascertain  that
$\chi(\mathbf{H}||z,\mathbf{l}||x)>\chi(\mathbf{H}||z,\mathbf{l}||y)$,
$\chi(\mathbf{H}||y,\mathbf{l}||x)>\chi(\mathbf{H}||z,\mathbf{l}||z)$
and
$\chi(\mathbf{H}||x,\mathbf{l}||y)>\chi(\mathbf{H}||x,\mathbf{l}||z)$.
I.e. the anisotropy of the transverse magnetic susceptibility due to
the Dzyaloshinskii-Moriya interaction favors the observed
spin-reorientation transitions.

\section{Conclusions.}

A detailed magnetic resonance study of the quasi-one-dimensional
antiferromagnet \bacusio{} is performed. We observed two low-energy
spin-resonance modes at  frequencies below 120GHz. Spin-
reorientation transitions are marked by the softening of one of the
resonance modes. Besides of the ordinary spin-flop transition a
spin-reorientation transitions with spins rotating in the plane
perpendicular to the magnetic field direction were identified.

The observed spin reorientations and magnetic resonance spectra in a
wide frequency range are described quantitatively in a model taking
into account the dependence  of the transverse susceptibility on the
direction of the order parameter. The role of this kind of anisotropy
is shown to be enhanced in a quasi one-dimensional antiferromagnet
due to the strong quantum reduction of the ordered spin component.

\acknowledgements

Authors thank M.E.Zhitomirsky (CEA-Grenoble, DRFMC/SPSMS) for his
continuous interest in this work and for fruitful discussions.
Authors are indebted to I.V.Telegina (Moscow State University,
Physics Department) for the invaluable help with the orientation of
single crystals.

The work was supported by the Russian Foundation for Basic Research
(RFBR), research project  03-02-16597 and by the INTAS project
99-0155.

\end{document}